\documentclass{svjour3}                     
\smartqed 

\usepackage{float}
\usepackage{graphicx}
\usepackage{pdflscape}
\usepackage{mathptmx}
\usepackage[numbers]{natbib}
\usepackage{hyperref}
\hypersetup{
	colorlinks	= true,
	linkcolor	= red,
	urlcolor	= cyan,
	citecolor	= blue
}

\newcommand{\exoclock}{\mbox{ExoClock}\ }

\begin{document}

\journalname{Experimental Astronomy}

\title{\exoclock Project: An open platform for monitoring the ephemerides of Ariel targets with contributions from the public
}

\titlerunning{\exoclock Project}

\author{
Anastasia Kokori$^{*1,2}$ \and
Angelos Tsiaras$^{1,2}$ \and 
Billy Edwards$^{1}$ \and
Marco Rocchetto$^{1,3}$ \and
Giovanna Tinetti$^{1}$ \and
Ana\"{e}l W\"{u}nsche$^{4}$ \and
Nikolaos Paschalis$^{5}$ \and
Vikrant Kumar Agnihotri$^{6}$ \and
Matthieu Bachschmidt$^{7}$ \and
Marc Bretton$^{4}$ \and
Hamish Caines$^{1}$ \and
Mauro Cal\'{o}$^{32}$ \and
Roland Casali$^{8}$ \and
Martin Crow$^{9, 10}$ \and
Simon Dawes$^{9, 10}$ \and
Marc Deldem$^{11}$ \and
Dimitrios Deligeorgopoulos$^{12}$ \and
Roger Dymock$^{9}$ \and
Phil Evans$^{13}$ \and
Carmelo Falco$^{14}$ \and
Stephane Ferratfiat$^{4}$ \and
Martin Fowler$^{9}$ \and
Stephen Futcher$^{15}$ \and
Pere Guerra$^{16}$ \and
Francois Hurter$^{17}$ \and
Adrian Jones$^{9}$ \and
Wonseok Kang$^{18}$ \and
Taewoo Kim$^{18}$ \and
Richard Lee$^{7}$ \and
Claudio Lopresti$^{19}$ \and
Antonio Marino$^{20}$ \and
Matthias Mallonn$^{21}$ \and
Fabio Mortari$^{22}$ \and
Mario Morvan$^{1}$ \and
Lorenzo V. Mugnai$^{33}$ \and
Alessandro Nastasi$^{14}$ \and
Valère Perroud$^{23}$ \and
Cédric Pereira$^{34}$ \and
Mark Phillips$^{9, 24}$ \and
Pavel Pintr$^{25}$ \and
Manfred Raetz$^{26}$ \and
Francois Regembal$^{27}$ \and
John Savage$^{9}$ \and
Danilo Sedita$^{28}$ \and
Nick Sioulas$^{29}$ \and
Iakovos Strikis$^{30}$ \and
Geoffrey Thurston$^{9}$ \and
Andrea Tomacelli$^{20}$ \and
Alberto Tomatis$^{31}$
}

\authorrunning{Anastasia Kokori et al.} 

\institute{
$^{*}$ anastasia.kokori@gmail.com
\\$^{1}$ University College London, Gower Street, London, WC1E 6BT, UK
\\$^{2}$ Aristotle University of Thessaloniki, University Campus, 54124 Thessaloniki, Greece
\\$^{3}$ Telescope Live, Spaceflux Ltd, 71-75 Shelton Street, London, WC2H 9JQ, UK
\\$^{4}$ Observatoire des Baronnies Proven\c cales, Le Mas des Gr\'es, Route de Nyons, 05150 Moydans, France
\\$^{5}$ Nunki Observatory, Skiathos, Greece (private observatory)
\\$^{6}$ Cepheid Observatory, Rawatbhata, India (private observatory)
\\$^{7}$ Amateur Astronomer
\\$^{8}$ Alto2000 Observatory, Italy (private observatory)
\\$^{9}$ British Astronomical Association, Burlington House, Piccadilly, Mayfair, London W1J 0DU
\\$^{10}$ Crayford Manor House Astronomical Society Dartford, Parsonage Lane Pavilion, Parsonage Lane, Sutton-at-Hone, Dartford, Kent, DA4 9HD
\\$^{11}$ Les Barres Observatory, Lamanon, France (private observatory)
\\$^{12}$ Artemis Observatory, Evrytania, Greece (private observatory)
\\$^{13}$ El Sauce Observatory, Coquimbo, Chile (private observatory)
\\$^{14}$ GAL Hassin - Centro Internazionale per le Scienze Astronomiche, Via della Fontana Mitri, 90010 Isnello, Palermo
\\$^{15}$ Hampshire Astronomical Group, Hinton Manor Ln, Clanfield, Waterlooville PO8 0QR, UK
\\$^{16}$ Observatori Astronòmic Albanyà, Spain (private observatory)
\\$^{17}$ Albireo Observatory, Switzerland (private observatory)
\\$^{18}$ National Youth Space Center, 11-1 Deokheung-ri, Dongil-myeon, Goheung-gun, Jeollanam-do, S. Korea
\\$^{19}$ GAD - Gruppo Astronomia Digitale, Italy
\\$^{20}$ Unione Astrofili Napoletani, Salita Moiariello, 16, CAP 80131 Napoli NA, Italy
\\$^{21}$ Leibniz Institute for Astrophysics Potsdam (AIP), An der Sternwarte 16, 14482 Potsdam, Germany
\\$^{22}$ Hypatia Observatory, Italy (private observatory)
\\$^{23}$ Observatoire de Duines, France (private observatory)
\\$^{24}$ Astronomical Society of Edinburgh, UK
\\$^{25}$ Institute of Plasma Physics AS CR, v. v. i., TOPTEC centre, Sobotecka 1660, 511 01 Turnov
\\$^{26}$ Herges-Hallenberg, Germany (private observatory)
\\$^{27}$ HRT Observatory, Canada (private observatory)
\\$^{28}$ Osservatorio Sedita Castrofilippo, Italy (private observatory)
\\$^{29}$ NOAK Observatory, Greece (private observatory)
\\$^{30}$ Hellenic Amateur Astronomy Association, Greece
\\$^{31}$ Alto-Observatory, Italy (private observatory)
\\$^{32}$ Cavallino Observatory, Tuscany, Italy (private observatory)
\\$^{33}$ La Sapienza Universita di Roma, Department of Physics, Piazzale Aldo Moro 2, 00185 Roma, Italy
\\$^{34}$ Instituto de Astrof\'isica e Ci\^encias do Espa\c co, Departamento de F\'isica, Faculdade de Ci\^encias, Universidade de Lisboa, Campo Grande, PT1749-016 Lisboa, Portugal
}

\date{Received: date / Accepted: date}

\maketitle

\begin{abstract}   

The Ariel mission will observe spectroscopically around 1000 exoplanets to further characterise their atmospheres. For the mission to be as efficient as possible, a good knowledge of the planets' ephemerides is needed before its launch in 2028. While ephemerides for some planets are being refined on a per-case basis, an organised effort to collectively verify or update them when necessary does not exist. In this study, we introduce the \exoclock project, an open, integrated and interactive platform with the purpose of producing a confirmed list of ephemerides for the planets that will be observed by Ariel. The project has been developed in a manner to make the best use of all available resources: observations reported in the literature, observations from space instruments and, mainly, observations from ground-based telescopes, including both professional and amateur observatories. To facilitate inexperienced observers and at the same time achieve homogeneity in the results, we created data collection and validation protocols, educational material and easy to use interfaces, open to everyone. \exoclock was launched in September 2019 and now counts over 140 participants from more than 15 countries around the world. In this release, we report the results of observations obtained until the 15h of April 2020 for 119 Ariel candidate targets. In total, 632 observations were used to either verify or update the ephemerides of 83 planets. Additionally, we developed the Exoplanet Characterisation Catalogue (ECC), a catalogue built in a consistent way to assist the ephemeris refinement process. So far, the collaborative open framework of the \exoclock project has proven to be highly efficient in coordinating scientific efforts involving diverse audiences. Therefore, we believe that it is a paradigm that can be applied in the future for other research purposes, too.

\keywords{Exoplanets \and Ephemerides \and Open Science \and Data analysis \and Photometry \and Citizen Science}

\end{abstract}

\section{Introduction}
\label{sec:introduction}

Just after discovery, the time of the next transit for a planet is well known. However, the accuracy of predicted future transits degrades over time due to the increased number of epochs since the last observation and the stacking of the period error \cite[e.g.][]{2019A_A...622A..81M, Dragomir2020, zellem2020}. In extreme cases this can mean the transit time is practically lost, with errors of several hours  \cite[e.g.][]{alonso}. In addition to this issue, extrapolating transit times from only a few data points over a limited baseline can easily introduce bias  \cite[e.g.][]{benneke, 2019A_A...622A..81M}. Finally, we could expect transit times to shift due to dynamical phenomena such as tidal orbital decay, apsidal precession or from gravitational interactions with other bodies in the system  \cite[e.g.][]{agol, maciejewski, bouma}. 

These issues can only be understood, and mitigated for, by regular observations over a long baseline and continuous ephemeris refinement. In this effort, small telescopes have been proven to be as efficient as larger ones \cite[e.g.][]{Beck2019, 2019A_A...622A..81M, Edwards2020, Kabath2019, zellem2020}. Especially for targets that are being discovered at the moment by TESS, estimations suggest that we will need efficient follow-up observations to verify their ephemerides and narrow down the large uncertainties in the predicted transit times \cite{Dragomir2020, zellem2020}. 

The Ariel mission \cite{Tinetti2018} will spectroscopically observe 1000 targets to study their atmospheres. To achieve this goal, thousands of transits will be observed within the lifetime of the mission. This large number underlines the necessity for precise predictions of the transit times, in order to maximise the overall efficiency of the mission. 

When planning observations for a single planet or for a small number of planets, ephemerides updates can be done on a per-case basis. However, in the new era of characterising large numbers of planets, such an effort needs to be organised in a much more efficient way through an open and collaborative platform, in order to make the best use of all the currently available resources. In several fields of astronomy there are public and collaborative databases which contribute significantly to community building and sharing. Examples of such efforts in astronomical fields are: VESPA\footnote{\href{http://www.europlanet-vespa.eu}{www.europlanet-vespa.eu}} \cite{Erard2018}, the Minor Planet Center\footnote{\href{https://www.minorplanetcenter.net/iau/mpc.html}{www.minorplanetcenter.net/iau/mpc.html}} and Simbad\footnote{\href{http://simbad.u-strasbg.fr/simbad}{simbad.u-strasbg.fr/simbad}}.

In this work, we present the \exoclock project (\href{https://www.exoclock.space}{www.exoclock.space}), an open, integrated platform with the scope of verifying the ephemerides of the planets that will be observed by Ariel, prior to the launch of the mission. \exoclock has been designed in a way to make the best use of all the available resources, including literature observations regardless of their purpose, and observations from any available ground- or space-based facilities, smaller or larger in size. In addition, \exoclock will provide a continuously updated and homogenous ephemerides service, which can be valuable for general exoplanet research, even beyond the Ariel mission.

\section{The \exoclock project}
\label{sec:exoclock}

Since Ariel will observe thousands of transits, good knowledge of the transit ephemerides is important in order to maximise the efficiency of the observing schedule and of the mission in general. The idea of the \exoclock project was conceptualised to achieve this scope, by delivering a verified list of ephemerides for the planets in the Ariel candidate target list  \cite{Edwards2019}, before the launch of Ariel. At the moment, this list includes about two thousand planets, from which 370 are already known and the rest are simulated future discoveries. From these planets, around 1000 will be chosen and observed by Ariel.

As the Ariel target candidate list includes thousands of planets, to achieve the goal of verifying their ephemerides we need to make the best use of all currently available resources and collect a large number of observations throughout many years. The \exoclock project has been designed to implement the above approach by:

\begin{enumerate}

\item integrating dedicated follow-up observations conducted by observers that have joined the project (forming the \exoclock network), literature observations, and observations conducted for other purposes, both from ground and space;
\item being open to contributions from a variety of audiences -- professional and amateur astronomers, students, and industry partners;
\item interacting with the participants through a variety of user-friendly interfaces to efficiently coordinate their efforts and ensure the high quality of the data products.

\end{enumerate}

The diversity of the project requires strategies to ensure homogeneity and reliability in the results. To achieve this vision, we have planned and applied several strategies which will be discussed in detail later. Since the project is open to contributions from inexperienced observers and at the same time contributions from various sources, it is very important to provide user-friendly interfaces and educational material. Throughout the course of the \exoclock project, we accomplish not only the most efficient use of resources but also a more vital connection between the several partners that collaborate. The open framework is an essential element in order to co-create knowledge effectively and share it with the rest of the community.\exoclock observations will be shared with the community every time a sufficient number of observations is collected.

\subsection{The citizen science aspects of \exoclock}
\label{sec:citizen_science}

Since \exoclock is open to contributions from everyone, inexperienced participants (e.g. amateur astronomers, students, citizen scientists, members of the public) constitute naturally an important part of the project. Currently, there are 140 participants in ExoClock, of which 80\% are amateur astronomers and the rest 20\% includes both early career and senior professional astronomers, but also a few school students and teachers. Therefore, the project has also a citizen science character. There are numerous citizen science projects designed to support scientific research and, indeed, citizen science is a mainstream trend. Citizen Science projects are efficient from various aspects \cite{Bonney2009}, such as:

\begin{itemize}
\item large-scale data collection;
\item elaborating scientific knowledge;
\item increasing the participation of a wider community;
\item strengthening the relationship between science and society;
\end{itemize}

However, several issues need to be considered as many of these projects have been questioned for their credibility and the accuracy in the delivered results. Hence, it is very important to plan and implement citizen science aspects carefully, considering a number of different parameters to ensure scientific robustness. In the \exoclock project we use different activities that are implemented at different stages during the project (data collection, data analysis and data validation). Following recommendations from successful citizen science projects \cite{Kosmala2016, Aceves-Bueno2017, Freitag2016}, we include strategies such as:

\begin{itemize}
\item educational material;
\item training sections;
\item interaction between the participants and the scientific team;
\item data collection protocols;
\item analysis tools;
\item validation to a reference database;
\item cross validation;
\item interactive project development;
\end{itemize}

\subsection{The Exoplanet Characterisation Catalogue (ECC)}
\label{sec:ecc}

As the final scientific product of this project is a verified catalogue of ephemerides, it was vital to construct an initial catalogue with the scope of continually updating it until Ariel launches. This catalogue was essential for the homogeneity in the final results but also for achieving the most efficient way of planning and analysing the observations collected by the participants.

The Exoplanet Characterisation Catalogue (ECC) includes the planets of the Ariel candidate target list that have already been discovered  \cite{Edwards2019}. At this moment, it includes 370 planets but it will be continuously updated based on upcoming discoveries. The catalogue contains stellar parameters (RA, DEC, magnitude at different filters, Temperature, Gravity and Metallicity) and also the transit parameters (ephemeris, period, Rp/ Rs, depth, inclination, transit duration and eccentricity). The ECC can be found as part of the data provided in this work (see Section \ref{sec:results}).

To ensure consistency in the catalogued values, we constructed the ECC following the rules below:
\begin{enumerate}
\item The target parameters (position, and magnitudes) were extracted from:
\begin{itemize}
\item SIMBAD for RA, DEC, and the magnitudes V, R, I
\item GAIA for the GAIA magnitudes G, G-BP, G-RP
\item 2MASS for the 2MASS magnitudes J, H, K
\end{itemize}
Where any of the V, R, I magnitudes were missing, we used the GAIA magnitudes to calculate them\footnote{\href{https://gea.esac.esa.int/archive/documentation/GDR2/Data\_processing/chap\_cu5pho/sec\_cu5pho\_calibr/ssec\_cu5pho\_PhotTransf.html\#Ch5.F15}{GAIA documentation}}.
This is indicated in the catalogue where applicable.
\item The stellar parameters used to calculate the limb darkening  - effective temperature, surface gravity, metallicity - were extracted from a single source for each planet (reference in the catalogue);
\item The transit parameters - planet radius, semi-major axis, eccentricity, inclination, argument of periastron - were extracted from a single source for each planet (reference in the catalogue);
\item The ephemeris parameters - zero-epoch mid-time, period - were extracted from a single source for each planet, the most recent one, but with the scope of updating them in the course of the project (reference in the catalogue).
\end{enumerate}

\subsection{Ephemeris verification criterion}
\label{sec:goal}

Observing before and after the transit is important to correct for instrumental systematics and Ariel will observe 75\% of the transit duration before and after each transit. Hence, we have defined that for an ephemeris to be considered as verified, we need to be sure that at least 50\% of the transit duration will be observed both before and after the transit. This is translated to a 3$\sigma$ uncertainty in the transit time prediction -- for the end of 2028 -- lower than 25\% of the transit duration, or to an 1$\sigma$ uncertainty lower than 1/12 of the transit duration (target uncertainty).

\section{Best use of resources}
\label{sec:resources}

In order to achieve the goals of the project it is fundamental to make the best use of all available resources. These include the observations available in the literature, and ongoing observations from space- and ground-based facilities, the latter including both professional and amateur observatories.

\subsection{The \exoclock network of telescopes}

In the \exoclock project we have developed an interface, where interested observers -- either professional or amateur astronomers -- can join the \exoclock network by registering their telescopes. So far, 140 observers have joined the \exoclock project, registering 180 telescopes to the \exoclock network. These telescopes are located in 15 countries around the world, and are of sizes between 6 and 40 inches.

Given the diversity in both the location and the sizes of the telescopes, we created a prioritisation system and a personalised scheduler for every registered telescope.

\subsubsection{Prioritisation system}
\label{sec:priorities}

We assigned an observing priority (high, medium, or low) to every planet in the ECC based on the 1$\sigma$ uncertainty in the transit time prediction (Eq. \ref{eq:uncertainty}):

\begin{equation}
\label{eq:uncertainty}
\sigma_T = \sqrt{\sigma_{T_0}^2 + (N\sigma_P)^2}
\end{equation}

where $T$ is the mid-time prediction, $T_0$ is the ephemeris mid-time, N is the number of epochs since the $T_0$, and $P$ is the orbital period of the planet.

We initially assigned an observing priority to every planet, as follows:

\begin{itemize}

\item{For planets with 1$\sigma$ uncertainty in the transit time prediction for the end of 2020 higher than 1/12 of the transit duration, the priority was set to high. Otherwise, if the ephemeris was older than four years, again the priority was set to high.}

\item{For the remaining planets, if the 1$\sigma$ uncertainty in the transit time prediction for the end of 2028 was higher than 1/12 of the transit duration, the priority was set to medium. Otherwise, if the ephemeris was between two and four years old, again the priority was set to medium.}

\item{For all the remaining planets, the priority was set to low.}
\end{itemize}

We also included a category of planets marked as ``alert", for which new observations suggest a time-shift higher than ten minutes. In this way we encourage observers to follow these specific planets until a clear picture about their ephemerides is formed.

During the course of the project the priorities are being dynamically updated based on the collected observations. For example, if a high priority planet is observed enough times during an observing season to improve or verify the ephemeris, its priority will be reduced to low. However, if the planet is not observed again in the following observing season, its priority will be increased again to medium.

\subsubsection{Personalised Schedule}
\label{sec:schedule}

To allocate more efficiently the targets to different observers, we have created a personalised schedule, which indicates to every observer the transits that are accessible to them, depending on their equipment and their location.

The expected signal-to-noise ratio is calculated based on Equation \ref{eq:snr} and if this is greater than 15, the planet is marked as observable.

\begin{equation}
\label{eq:snr}
SNR = a D \sqrt{10 ^ {\frac{12 - R_{mag}}{2.5}}} \frac{T_{Dp}}{ \sqrt{ \frac{1}{T_{Dr}} + \frac{1}{120}} }   
\end{equation}

where $a$ is a constant, $D$ is the telescope aperture in inches, $R_{mag}$ is the R magnitude of the star, $T_{Dp}$ is the transit depth in mmag, and $T_{Dr}$ is the transit duration in minutes. 

The constant $a$ was initially calculated based on observations we acquired from the Holomon Astronomical station of the Aristotle University of Thessaloniki in Greece. They represent a system with an 11 inch aperture, an ATIK11000 camera and a Red (Cousins) filter, which is widely considered as a typical equipment capable of conducting transit observations. For this system and assuming an observation that starts one hour before the transit, ends one hour after the transit, has an exposure time of one minute, and has overheads of 30 seconds, $a$ is equal to 0.0125.

As observers submit their observations, we track the performance of their systems and further adjust the $a$ coefficient per system. In this way the schedule takes into account the capabilities of the different instruments and the long-term effects of weather patterns, in order to distribute the planets accordingly. In terms of timing, only the observations that happen when the star is at least 20 degrees above the horizon appear in the scheduler. Finally, we make sure that the Sun stays below -10 degrees throughout the observation.

Based on the above calculation, and placing an upper limit for the transit duration to 6 hours, we computed the number of planets in the Ariel candidate target list that are accessible through the ExoClock network. For telescopes of 10, 15, 25, 35, and 40 inches the percentages are 14\%, 22\%, 36\%, 47\%, 52\%, respectively. If we however relax the requirement of SNR=15 to SNR=10, these numbers increase to 22\%, 33\%, 49\%, 62\%, 67\%, respectively.

\subsection{Space-based facilities}
\label{sec:space}

Space-based follow up will be needed for targets that are not accessible from the ground (of low S/N or very long duration). ExoClock will integrate past and future results from the main space telescopes that have observed or will observe planets in the ECC.

\subsubsection{TESS}
\label{sec:tess}

During its primary mission, TESS is surveying almost the entire sky. While the prime objective of the mission is to discover new planets, TESS will also observe stars which are known to host planets. As of November 2019, TESS has observed the host stars of 150 planets within the Ariel target list. However, data of sufficient quality has not necessarily been obtained for each of these. Thus, while TESS will undoubtedly be a great resource for the refinement of transit times  \cite[e.g.][]{Bouma2019, Edwards2020, Smith2020}, the extent of this contribution needs further exploration.

\subsubsection{Hubble and Spitzer}
\label{sec:tess}

Spitzer has been used to follow-up several exoplanets with ephemeris uncertainties \cite[e.g.][]{benneke, livingston, kosiarek} and the data will be valuable for future ephemeris refinements. However, the spacecraft has ceased operations in early 2020 and no further observations are available. The Hubble Space Telescope (HST) has been delivering spectroscopic observations of exoplanets since 2001. Although most observations are interrupted due to Earth obscuration, the precision can still be sufficient for measuring accurate transit times  \cite[e.g.][]{Skaf2020, Bruno2018}. HST may well be used to observe new TESS detections and could therefore provide ephemeris refinement. However, HST cannot be used in cases where the uncertainty on the transit time is too large. Also, HST is a general observatory and can only observe a limited number of targets. Therefore, neither of these missions can be expected to have a major impact on any ephemeris refinement project.

\subsubsection{CHEOPS}
\label{sec:tess}

CHEOPS launched in December 2019 and is anticipated to follow-up a large number of TESS discoveries, refining both planetary and orbital parameters via high precision photometry  \cite{Broeg2013}. The mission has reserved around 20\% of the telescope time for guest observers which equates to 946 orbits, or 1578 hours, in the first year. To allow potential users to explore the capabilities of CHEOPS, a simulator has been developed which has a web interface\footnote{\href{https://cheops.unige.ch/pht2/exposure-time-calculator}{https://cheops.unige.ch/pht2/exposure-time-calculator}}. 

By inputting the star type and magnitude, along with the observation duration, a noise prediction is returned. The web-based simulator has been utilised to understand the performance for ephemeris refinement of planets within the Ariel target list. CHEOPS cannot observe the ecliptic poles, which are the continuous viewing zones for JWST and Ariel, and planets outside the field of regard have been removed. From this exercise, we conclude that additional 300 planets (15\%) that are accessible to CHEOPS with an SNR above 5 in one observation. Further investigation is needed to ascertain the exact number of targets for which CHEOPS could reduce the transit uncertainties but it would seem likely that the final figure will be of the order of several hundred.

\subsection{Other ground-based networks}
\label{sec:ground}

\exoclock as an integrated platform will incorporate all available observations produced by other ground-based networks, too. We are already collaborating with other networks and current projects to both make the best use of all resources and decrease the potential for overlapping efforts. Since homogeneity is key for the \exoclock project, we are currently examining the data sharing protocols and quality control protocols. Data from such databases will be included in the \exoclock database in future releases after defining the most effective strategy to collect them ensuring validity and accuracy. 

Although such networks carry out transit exoplanet observations, they are focused on different aspects, as far as exoplanet ephemerides are concerned. For example, the Exoplanet Transit Database \cite[ETD,][]{Poddany2010} run by the Czech Astronomical Society since 2009, provides more than 10,000 transit light curves for more than 350 exoplanets systems, with the scope of searching for transit time variations. Additionally, Exoplanet Watch \cite{zellem2020} is another coordinated effort to collect follow-up observations of exoplanet transits with small telescopes, organised in the USA and focused on targets that will be then observed by the James Webb. Space Telescope. 

Finally, the ExoClock project is in close collaboration with the Telescope Live\footnote{\href{https://telescope.live}{https://telescope.live}} network. Telescope Live is a web application offering end-users the possibility to purchase images obtained on-demand from a network of robotic telescopes. At the time of writing the network consists of eight telescopes distributed across three observatories: a 1m ASA RC-1000AZ, a 0.6m Planewave CDK24 and two 0.5m ASA 500N located at El Sauce Observatory in Chile; a 0.7m ProRC 700 and two 0.1m Takahashi FSQ-106ED located at IC Astronomy in Spain; a 0.1m Takahashi FSQ-106ED located at Heaven’s Mirror Observatory in Australia.

\subsection{Catalogue of literature observations}
\label{sec:literature}

Since there is currently no organised and easily accessible database of past observations, conducted by other observatories (ground- or space-based) and recorded in the literature, it is difficult for the scientific community to distinguish which planets need their ephemerides to be updated.

To facilitate easier access to the current literature we have developed an application within the website, in which participants and members of the scientific community can add results from other published works. The application includes all the appropriate conversions between the different time systems used across the different papers to the BJD$_\mathrm{TDB}$ system.

These contributions are validated and they are taken into account in the ephemerides refinement. In this work we include the literature data only for the planets for which an ephemeris update is required. We will continue developing this platform and collect more data from past and new publications, to ensure the best use of all transit observations, even beyond the \exoclock platform.

\section{Scientific robustness and homogeneity of the results}
\label{sec:homogeneity}

Ensuring homogeneity and maintaining high quality in the results produced is crucial when multiple observers are participating. To achieve this requirement, it is mandatory to base the analysis on a single catalogue, and this was one of the main reasons for constructing the ECC described above. In addition, we have established a protocol for conducting and analysing observations, and specific educational material to support the observers.

\subsection{Observing strategy}
\label{sec:observing}

Based on experience from previous observations and collaboration with other experts on exoplanets research, we created an observing strategy to be followed as a suggestion by all our participants (data collection protocol). In this strategy we pay extra attention to ensure the observability of the transits, to maximise the total S/N ratio of the light curves and to achieve the highest possible precision in transit timing. More specifically, the observers are recommended to:

\begin{itemize}

\item follow the personalised schedule to only follow-up planets that can be observed with their equipment and from their location. (see Section \ref{sec:schedule} for more details);
\item use an R or I Cousins filter, preferably. Alternatively, a luminance or clear filter is suggested. In general, the precision in the measurement of the mid-time of a transit increases with wavelength. The reasons for this improvement are the weaker stellar activity, the weaker limb darkening effect and the lower background noise;
\item set the camera temperature to the lowest available value. This step is necessary to decrease the noise from dark current as much as possible;
\item set the binning to one to allow longer exposure times, decreasing in this way the total overheads and increasing the total S/N;
\item keep the number of counts below the point where the camera becomes non-linear. This procedure is suggested to avoid biases in the transit depth caused by the non-linear response of the camera;
\item check for at least one good comparison star in the field and adjust it if it is needed;
\item observe for an extra hour before the start of the transit and an extra hour after the end of the transit. This strategy is necessary to model efficiently any instrumental systematics (see Section \ref{sec:fitting}) for details on the systematics modelling). In addition, it helps capturing unexpected drifts between the predicted and the observed transit mid-times;
\item validate the clock of their computer against the network time. A precision of one to two seconds can be easily achieved in this way, and it is enough for the purposes of this project.

\end{itemize}

\subsection{Data reduction and photometry}
\label{sec:reduction}

The data collection protocol described above is supported by a dedicated piece of software for data reduction and photometry. We specifically developed this software (the HOlomon Photometric Software\footnote{\href{https://github.com/ExoWorldsSpies/hops}{github.com/ExoWorldsSpies/hops}}) to certify both the high quality and the homogeneity of the results but also to ensure accessibility for users of different experience levels, starting from users who had never observed transits before.

The software is developed in Python, allowing community development and easy adaptation to new features. It is also continuously developed in collaboration with amateurs who provide feedback on the user experience and the performance of the software under different circumstances. HOPS supports the users by being user friendly; interactive, with user interfaces; open source; free; geospacial; and compatible with all operational systems.

As far as the data analysis is concerned, HOPS allows the user to reduce and align the data; to select the best comparison stars; to extract the target's light curve; and to perform an MCMC fitting to allow for a first validation of the results. Providing more technical details is beyond the scope of this work and will be discussed in a separate, dedicated, publication.

\subsection{Light curve fitting}
\label{sec:fitting}

While the data reduction and photometry can be performed using the observer's preferred tools, it is of great importance to perform the light curve modelling on the \exoclock web-server. We use the open source Python package PyLightcurve\footnote{\href{https://github.com/ucl-exoplanets/pylightcurve}{github.com/ucl-exoplanets/pylightcurve}} \cite{Tsiaras2016B2016ascl.soft12018T} to facilitate this process.

At first, the time basis of the uploaded data is converted to BJD$_\mathrm{TDB}$. This step is necessary as the observers are allowed to upload their data in any available time basis. However, we recommend JD$_\mathrm{UTC}$ to be used, as the conversions applied by different software may be subject to assumptions that are not reversible. Also, the data are converted from magnitude to relative flux, if necessary. If the uploaded data do not include uncertainties, these are estimated with a moving standard deviation. In both cases, we perform a first fitting, and then we scale the uncertainties to match the standard deviation of the first fitting residuals. Then, the fitting process is repeated. This is a standard process in the literature applied to observations of HST \cite{Tsiaras2018}. In this way, we take into account any extra scatter in the observation and end up with a conservative estimation of the uncertainties in the final results.

The limb darkening coefficients are calculated based on the stellar parameters (temperature, gravity and metallicity) for the planet as provided in the ECC, the Phoenix 3D stellar models and the response curves of the different filters, through the open source Python package ExoTETHyS\footnote{\href{https://github.com/ucl-exoplanets/ExoTETHyS}{github.com/ucl-exoplanets/ExoTETHyS}} \cite{Morello2020}. For all the models, we use the 4-coefficients law by Claret 2000 \cite{Claret2000}.

During the fitting, we fix all the transit parameters, with the exception of the planet-to-star radius ratio and the mid-time, to the values in the ECC. Simultaneously with the transit model, we use a model for the out-of-transit systematics, which are usually due to airmass variations. We consider three models for the systematics (linear with airmass, linear with time, and quadratic with time) and choose the one which best describes the data (minimum reduced chi-square and residuals autocorrelation), at the same time, giving a result for the planet-to-star radius ratio close to the expected value from ECC. We need to note here that all the models used are exposure-time integrated -- i.e. they do take into account the variation in the light curve during the exposure. This is achieved by calculating the model for sub-exposures of 10 seconds and then using the average.

The results from each individual light curve fitting are compared with both the literature and other observations on \exoclock. Based on the information from these sources, each observation is either validated by the platform, or returned to the user for further reduction and photometry.

\subsection{Educational material}
\label{sec:1}

Continuous training and interaction between the participants and the science team are fundamental elements to ensure high quality in the produced results. The learning process is complex and multi-dimensional and every person learns in a different way. For these reasons, it is important to support the learning process through a variety of material: manuals; guidelines; videos; podcasts; newsletters; one-to-one feedback; training datasets.

Because our project is open to people from various countries, it is significant to facilitate their participation and avoid misconceptions. Thus, a large part of the educational material is also provided in different languages (at the moment English, Greek, Portuguese, and French) and we also have national contact points for several countries, to assist at a local level. All this material is hosted on a separate website, especially built for educational purposes\footnote{\href{https://www.exoworldsspies.com}{www.exoworldsspies.com}}.

\section{Results - first data release}
\label{sec:results}

Here we present the first data release of the \exoclock project, which includes three data products:

\begin{itemize}
\item the Exoplanet Characterisation Catalogue (see Section \ref{sec:ecc})
\item the catalogue of \exoclock observations
\item the catalogue of \exoclock ephemerides
\end{itemize}

All data products and their descriptions can found through the OSF repository with DOI: \href{http://doi.org/10.17605/OSF.IO/3W7HM}{10.17605/OSF.IO/3W7HM}

\subsection{Catalogue of \exoclock observations}
Here we present the first data release of the \exoclock project, containing 632 observations of 120 planets in the ECC (30\%). These observations were conducted between 2011 and 2020, submitted to the \exoclock platform before the 15$^\mathrm{th}$ of April 2020, and validated according to the criteria described in Section \ref{sec:fitting}. The current rate of submitting new observations is above 100 per month. From the 632 observations in this dataset, 402 (17\% from amateur astronomers) were acquired before the kick off of the project in September 2019, while the remaining 230 (73\% from amateur astronomers) after September 2019. The largest contribution to this dataset has been made by the Observatoire des Baronnies Proven\c cales with 356 observations, most of which were acquired before September 2019.

\begin{figure}
\centering
\includegraphics[width=\columnwidth]{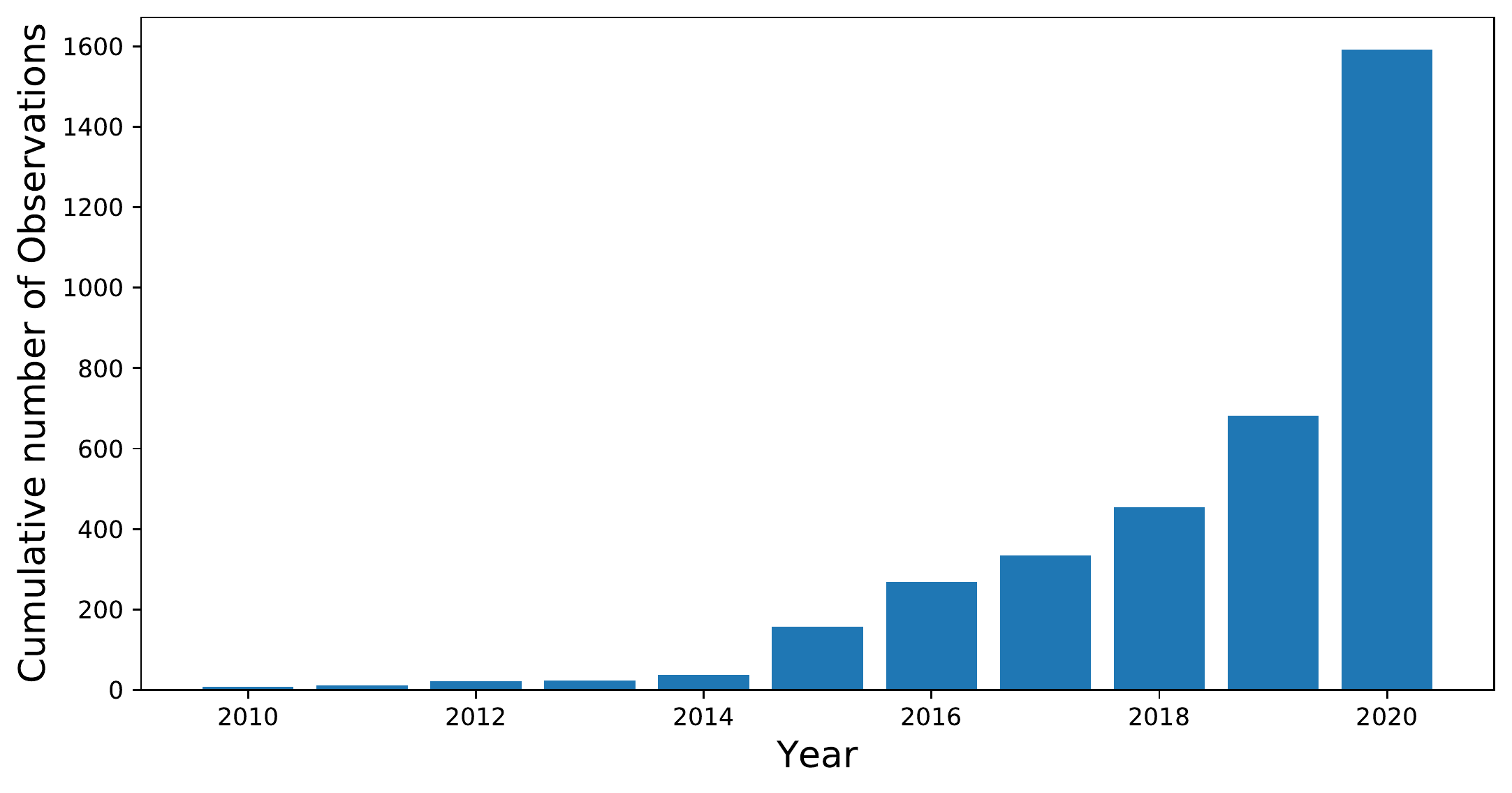}
\caption{Cumulative distribution of all the observations in the \exoclock platform, based on the observing date.}
\label{fig:observations}
\end{figure}

Note that in this work, only the observations that were submitted up to the 15$^\mathrm{th}$ of April 2020 are included. At the moment, the \exoclock platform contains more than 1500 observations (62\% from amateur astronomers), which will be analysed in future publications. In addition, the ECC will be dynamically updated in the future, as more planets that are accessible by Ariel will be discovered. Figure \ref{fig:observations} shows the cumulative distribution of all the observations based on the time of observation (not the time of submission).

The online catalogue of ExoClock observations includes the following information for each observation:
\begin{enumerate}
\item the raw light curve;
\item metadata regarding the observer(s), the planet observed (link to ECC), the equipment used (telescope-camera-filter), the exposure time and the time and flux formats;
\item the raw light curve filtered for outliers, converted to BJD$_\mathrm{TDB}$ and flux formats and enhanced with the estimation for the uncertainties, the target altitude, and the airmass;
\item the fitting results, including the de-trending method used and its parameters;
\item the de-trended light curve, enhanced with the de-trending model, the transit model and the residuals;
\item fitting diagnostics on the residuals (standard deviation, chi-squared, autocorrelation)
\end{enumerate}

\subsection{Catalogue of \exoclock ephemerides}

The \exoclock project aims to deliver a catalogue of verified ephemerides for the planets in the Ariel target list, before the launch of the mission. An ephemeris is considered verified if it gives predictions for the end of 2028 with an uncertainty lower than 1/12 of the transit duration (see section \ref{sec:goal}). This catalogue is initiated here with verified ephemerides for 84 planets in the ECC (22\%). The verified status of these ephemerides is clearly indicated on the website, to encourage their use by the community. The 84 ephemerides that have been verified will be continuously monitored, as biases in measurements may still exist but require more observations to become statistically significant, and their status will be revised with future data releases.

To produce this catalogue we assessed the ephemerides of the 120 planets observed in this first data release as follows:

\begin{enumerate}
\item We updated the ephemerides that did not meet the verification criterion using both the \exoclock observations and values from the literature (19 in total). Note that for some of these planets, data existed in the literature but the ephemerides were either not updated by the authors or the updates where not properly listed in other catalogues. This proves the need for an organised way of listing the ephemeris information for all planets.

\item We evaluated the O-C drift for the remaining ephemerides and identified those for which our observations suggested an important, unexpected, drift (9 in total). An important drift is one that predicts an O-C greater than 1/12 of the transit duration for the end of 2028. We then updated these ephemerides using both the \exoclock observations and values from the literature. Note that these planets met the verification criterion, and the drifts observed were due to biases in the initial ephemerides. This proves the need for continuous and regular follow-up after the transit discovery.

\item
We verified the current ephemerides for the planets that were observed at least twice and did not show any significant drift (56 in total). The majority of the planets in this category were marked as low priority.

\item The remaining 36 ephemerides were not verified with the data in this release as only one observation was acquired.\end{enumerate}

For planets in categories 1 and 2, where the ephemerides updates were vital for continuing monitoring the planets in the future, we did use single observations in some cases (KELT-15b, WASP-13b, WASP-31b, WASP-67b, KELT-7b, WASP-16b). In these cases, there was the need for the update and also there were data, either in the literature or on our website, supporting the observed drift. To update the ephemerides we considered both literature mid-times and the mid-times that resulted from analysing the ExoClock observations. At first, we computed a new zero-epoch as the weighed average of the epochs of the available data.

\begin{figure}[H]
\centering
\includegraphics[width=\columnwidth]{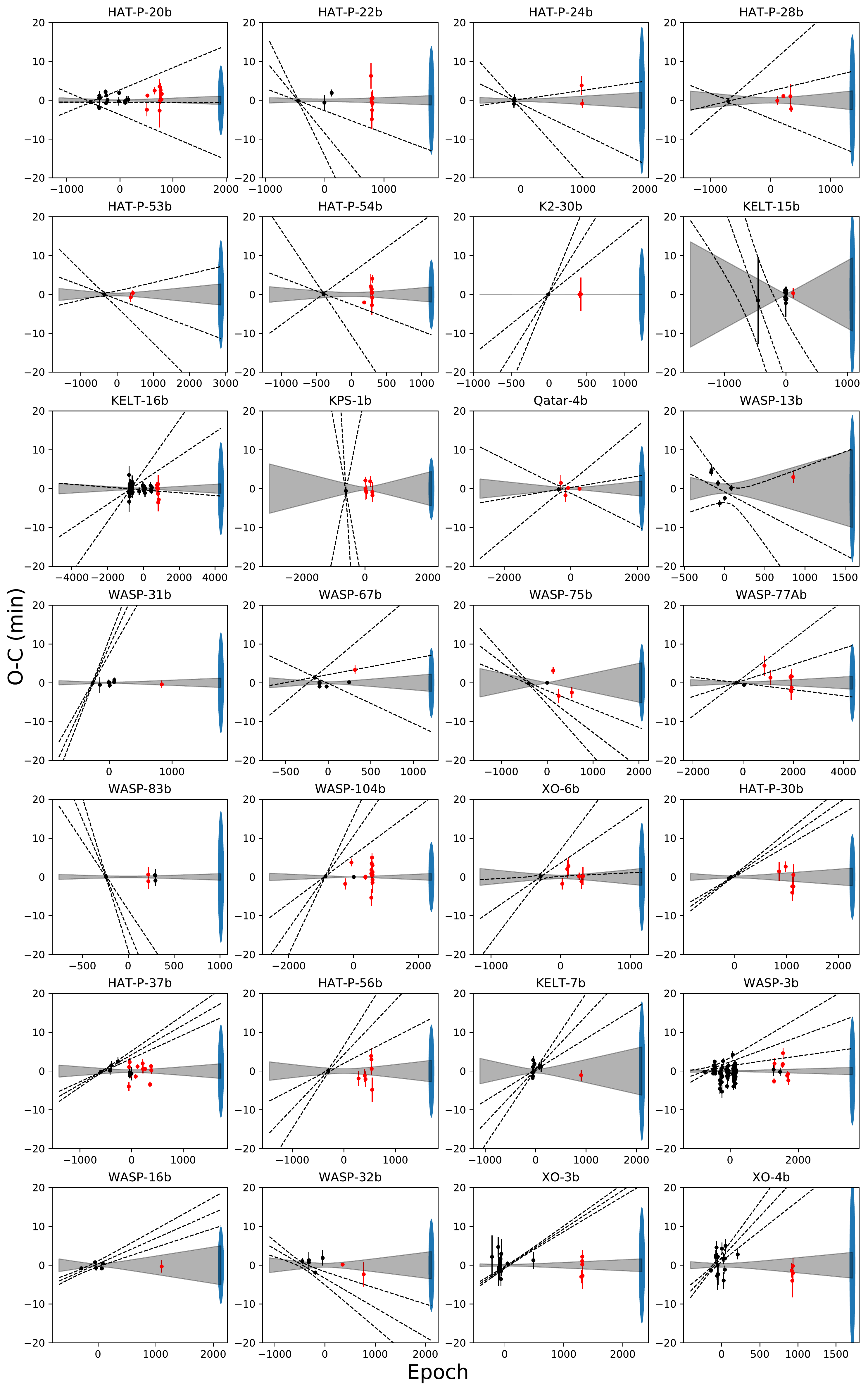}
\caption{Comparison between the initial (dashed lines) and the updated ephemerides (shaded area) for the 28 planets that had their ephemerides refined. The black circles indicate literature values while the red ones indicating observations provided by telescopes registered in the \exoclock network. In blue, we indicate the target 1$\sigma$ uncertainty for the end of 2028 for each planet. For all planets, the horizontal axis represents epochs between  01/01/2005 and 01/01/2029.}
\label{fig:ephemerides}
\end{figure}

\begin{figure}[H]
\centering
\includegraphics[width=\columnwidth]{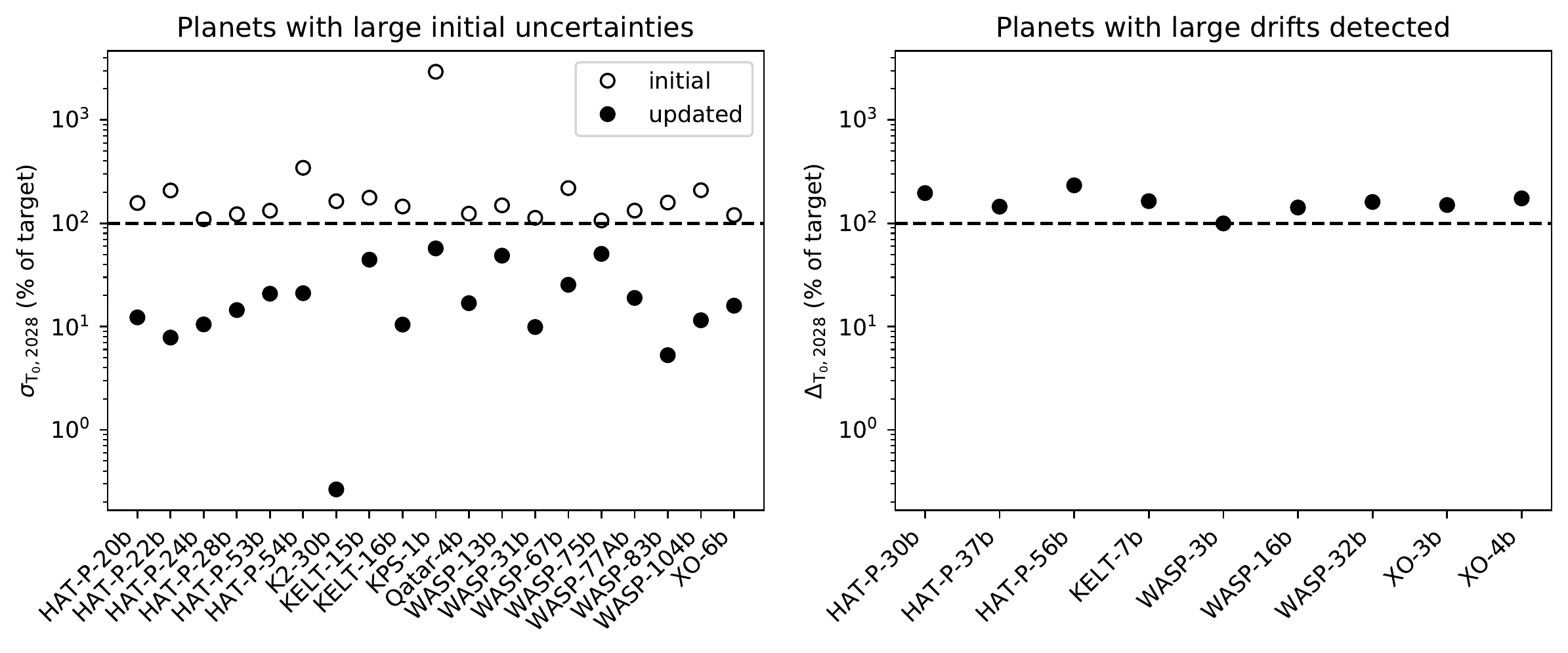}
\caption{Left: Comparison between the initial and the updated uncertainties in predictions for the end of 2028, relatively to the target uncertainty (1/12 of the transit duration) for updated planets in the category 1. Right: Drift of the mid-time predictions for the end of 2028 between the initial and the updated ephemerides for those planets which were found to have unexpected drifts.}
\label{fig:drifts}
\end{figure}

For the 28 ephemerides that have been updated, we refer the reader to the online data repository for all the data used for the refinement. Figure \ref{fig:ephemerides} show the all the 28 refined ephemerides compared with the initial ones while, Figure \ref{fig:drifts} shows the variations in the precision and the 2028 predictions for the refined ephemerides. In addition,
Table \ref{tab:updated} presents the ephemerides that have been updated (categories 1 and 2), Table \ref{tab:verified} presents the ephemerides that have been verified (category 3), and finally Table \ref{tab:ref} gives the references for the parameters used in the analysis of all the planets in this data release.

\section{Discussion}
\label{discussion}

The results presented here demonstrate that a manifold platform like \exoclock is very efficient in making the best use of all available resources towards addressing a defined scientific problem. In our case, the defined problem was the need for up-to-date ephemerides in order to increase the efficiency of the Ariel observations and other follow-up activities. However, the project architecture (priorities, organisation, website features etc) could support other endeavours, either in the same field or other sectors, to optimise data collection and ensure reliability and homogeneity in the results.

The open nature of the \exoclock project guarantees the inclusion of various audiences that have the opportunity to contribute directly. This is fundamental to achieve the best use of all currently available resources. Currently, there are 140 participants in \exoclock, of which 80\% are amateur astronomers and the rest include both early career and senior professional astronomers, but also a few students and teachers. These various groups of people come from different backgrounds and have different interests. This is significant since it results in a dynamic team of people who continuously bring new ideas and enrich the complexion of the project. Since also \exoclock observers have different levels of experience, they contribute to the project from different perspectives. 

Through the course of the project, it became clear that an integrated platform like \exoclock is beneficial to the whole exoplanet community as it can support maximising the efficiency of any organised observing effort and maintaining an always up-to-date scientific product. This is done through providing a uniformly derived reference catalogue, allocating observations according to the capabilities of the observatories (personal schedule), and prioritising the observations. 

One of the most important elements in the \exoclock platform is the alert system, which was developed especially for planets that showed unexpected drifts. This system proved very effective so far, with the participants being very responsive to such alerts. Apart from large O-C drifts, the alert system helped us identify mistakes in the initial catalogues, related to either the ephemerides or the transit parameters. An example was the case of WASP-84b, for which the initial ephemeris was mistyped in the NASA Exoplanet archive, leading to wrong transit predictions by many hours. Similarly, a mistyped value for Kepler-447b led to a much larger transit depth, a mistake identified after observations made by the \exoclock participants. These results would not had be identified and cross-validated without the participation of an active community that is always willing to observe, even planets with low or medium priority.

Certainly, the highly interactive design of the \exoclock platform is another key element, and it is achieved through various approaches, such as: educational material, training, personal feedback, meeting, newsletters, and, even more importantly, the development of user friendly tools that enable participation from inexperienced observers. This is an element that contributes significantly towards achieving homogeneity in the results, as it supports the efficient implementation of the data collection and analysis protocols described above. It is important to note that 70\% of the \exoclock participants use HOPS for data reduction and photometry.

\section{Conclusions}
\label{conclusions}

From the experience gained in \exoclock, we identified the importance of making the best use of all available resources through organising community-wide efforts. This is a prototype that can be used in other fields apart from exoplanets. Organising a project in a collaborative perspective by considering contributions from various audiences can maximise the outcome of the available resources. To implement efficiently such a project it is vital to create data collection and validation protocols, educational material (guides, newsletters) and easy-to-use interfaces (user-friendly software and website). In future studies, we will provide the ephemerides verification and updates for more Ariel targets according to new discoveries. The monitoring of the Ariel candidate targets will be continued until the launch of the mission but \exoclock can be established as a platform and used by the exoplanet communities for further purposes. The open framework of \exoclock is vital to facilitate future exoplanet research and avoid wasting resources.

\section*{Software and Data} 

Software used: Django, PyLightcurve \cite{Tsiaras2016B2016ascl.soft12018T}, ExoTETHyS \cite{Morello2020}, Astropy \cite{astropy}, emcee \cite{emcee}, Matplotlib \cite{matplotlib}, Numpy \cite{numpy}, SciPy \cite{scipy}.

All the data products and their descriptions can found through the OSF repository with DOI: \href{http://doi.org/10.17605/OSF.IO/3W7HM}{10.17605/OSF.IO/3W7HM}.

\section*{Acknowledgements}
This project has received funding from the European Research Council (ERC) under the European Union's Horizon 2020 research and innovation programme (grant agreement No 758892, ExoAI). Furthermore, we acknowledge funding by the Science and Technology Funding Council (STFC) grants: ST/K502406/1, ST/P000282/1, ST/P002153/1, and ST/S002634/1. L.V. Mugnai is supported by ASI grant n. 2018.22.HH.O. P. Pintr acknowledges the support by the project No.CZ.02.1.01/0.0/0.0/16\_026/0008390.

\section*{Conflict of interest}
The authors declare that they have no conflict of interest.

\newpage

\newpage

\begin{table}[H]
\centering
\caption{New ephemerides for the 28 planets that have been updated here (categories 1 and 2), and references for the literature data used.}
\label{tab:updated}
\begin{tabular}{lccl}
Planet & New T$_0$ (BJD$_\mathrm{TDB}$) & New Period (days) & References for \\
       &                                &                   & literature data used  \\ [0.3ex] \hline

HAT-P-20b & 2456670.97803 $\pm$ 0.0001 & 2.875317 $\pm$ 4e-07 & \cite{2011ApJ...742..116B} \cite{2017AJ....153...28S} \cite{2014AN....335..797G} \cite{2017A_A...601A..53E} \\ [0.3ex] 
HAT-P-22b & 2456366.08903 $\pm$ 0.00025 & 3.212233 $\pm$ 4e-07 & \cite{2011ApJ...742..116B} \cite{2016MNRAS.459..789T} \\ [0.3ex] 
HAT-P-24b & 2455592.76491 $\pm$ 0.00023 & 3.3552454 $\pm$ 7e-07 & \cite{2010ApJ...725.2017K} \\ [0.3ex] 
HAT-P-28b & 2457713.934 $\pm$ 0.0005 & 3.2572124 $\pm$ 1.2e-06 & \cite{2011ApJ...733..116B} \\ [0.3ex] 
HAT-P-53b & 2456514.0562 $\pm$ 0.00025 & 1.9616265 $\pm$ 7e-07 & \cite{2015AJ....150..168H} \\ [0.3ex] 
HAT-P-54b & 2457811.6454 $\pm$ 0.0004 & 3.7998518 $\pm$ 1.1e-06 & \cite{2015AJ....149..149B} \\ [0.3ex] 
K2-30b & 2457112.988874 $\pm$ 1.3e-06 & 4.098481163 $\pm$ 1.8e-08 & \cite{2016AJ....151..171J} \\ [0.3ex] 
KELT-15b & 2458544.07707 $\pm$ 0.00018 & 3.329472 $\pm$ 6e-06 & \cite{2016AJ....151..138R} \cite{2020MNRAS.tmp.1386E} \\ [0.3ex] 
KELT-16b & 2457946.86088 $\pm$ 0.0001 & 0.96899295 $\pm$ 2e-07 & \cite{2017AJ....153...97O} \cite{2018AcA....68..371M} \cite{2020AJ....159..150P} \\ [0.3ex] 
KPS-1b & 2458544.108 $\pm$ 0.0003 & 1.7063219 $\pm$ 1.5e-06 & \cite{2018PASP..130g4401B} \\ [0.3ex] 
Qatar-4b & 2458293.1227 $\pm$ 0.00015 & 1.805369 $\pm$ 6e-07 & \cite{2017AJ....153..200A} \\ [0.3ex] 
WASP-13b & 2455218.5681 $\pm$ 0.0008 & 4.353019 $\pm$ 4e-06 & \cite{2009A_A...502..391S} \cite{2012MNRAS.419.1248B} \\ [0.3ex] 
WASP-31b & 2456088.4374 $\pm$ 0.0001 & 3.4058853 $\pm$ 5e-07 & \cite{2011A_A...531A..60A} \cite{2011AJ....142..115D} \cite{2015MNRAS.446.2428S} \\ [0.3ex] 
WASP-67b & 2456548.83747 $\pm$ 0.0002 & 4.6144172 $\pm$ 1.3e-06 & \cite{2012MNRAS.426..739H} \cite{2014A_A...568A.127M} \cite{2018AJ....155...55B} \\ [0.3ex] 
WASP-75b & 2457007.46252 $\pm$ 5e-05 & 2.4841993 $\pm$ 1.7e-06 & \cite{2013A_A...559A..36G} \cite{2018PASP..130c4401C} \\ [0.3ex] 
WASP-77Ab & 2456224.05817 $\pm$ 0.00017 & 1.3600295 $\pm$ 3e-07 & \cite{2013PASP..125...48M} \cite{2016MNRAS.459..789T} \\ [0.3ex] 
WASP-83b & 2457121.99609 $\pm$ 0.00015 & 4.9712917 $\pm$ 6e-07 & \cite{2015AJ....150...18H} \cite{2020MNRAS.tmp.1386E} \\ [0.3ex] 
WASP-104b & 2457928.048611 $\pm$ 1.7e-05 & 1.7554057 $\pm$ 3e-07 & \cite{2014A_A...570A..64S} \cite{2018AJ....156...44M} \\ [0.3ex] 
XO-6b & 2457737.031 $\pm$ 0.0003 & 3.7649924 $\pm$ 1.3e-06 & \cite{2017AJ....153...94C} \\ [0.3ex] 
HAT-P-30b & 2455765.63288 $\pm$ 0.00018 & 2.8106019 $\pm$ 7e-07 & \cite{2011ApJ...735...24J} \cite{2011AJ....142...86E} \cite{2016AcA....66...55M} \\ [0.3ex] 
HAT-P-37b & 2457303.82448 $\pm$ 0.00024 & 2.7974418 $\pm$ 7e-07 & \cite{2012AJ....144...19B} \cite{2016AcA....66...55M} \cite{2017MNRAS.472.3871T} \\ [0.3ex] 
HAT-P-56b & 2457404.8183 $\pm$ 0.0004 & 2.7908233 $\pm$ 1.1e-06 & \cite{2015AJ....150...85H} \\ [0.3ex] 
KELT-7b & 2456385.3121 $\pm$ 0.0004 & 2.7347694 $\pm$ 2e-06 & \cite{2015AJ....150...12B} \\ [0.3ex] 
WASP-3b & 2455517.89652 $\pm$ 0.0001 & 1.84683506 $\pm$ 1.8e-07 & \cite{2008MNRAS.385.1576P} \cite{2010ApJ...715..421T} \cite{2008A_A...492..603G} \cite{2011ApJ...726...94C} \cite{2013A_A...549A..30N} \\ & & & \cite{2012PASP..124..212S} \cite{2012MNRAS.427.2757M} \cite{2010MNRAS.407.2625M} \cite{2012MNRAS.423.1381E} \cite{2013AJ....146..147M} \\ & & & \cite{2018IBVS.6243....1M} \\ [0.3ex]
WASP-16b & 2455495.0619 $\pm$ 0.0003 & 3.1186023 $\pm$ 1.6e-06 & \cite{2009ApJ...703..752L} \cite{2013MNRAS.434.1300S} \\ [0.3ex] 
WASP-32b & 2456379.8909 $\pm$ 0.0004 & 2.7186643 $\pm$ 1.1e-06 & \cite{2010PASP..122.1465M} \cite{2015RAA....15..117S} \cite{2012PASP..124..212S} \\ [0.3ex] 
XO-3b & 2454721.14865 $\pm$ 0.00014 & 3.191526 $\pm$ 5e-07 & \cite{2008ApJ...677..657J} \cite{2008ApJ...683.1076W} \cite{2017AN....338...35G} \cite{2008A_A...488..763H} \cite{2009ApJ...700..302W} \\ & & & \cite{2017MNRAS.472.3871T} \\ [0.3ex]
XO-4b & 2455067.5685 $\pm$ 0.0003 & 4.1250683 $\pm$ 1.3e-06 & \cite{2008arXiv0805.2921M} \cite{2010PASJ...62L..61N} \cite{2016ApJ...820...87V} \cite{2012ApJ...746..111T} \\ [0.3ex] 

\end{tabular}
\end{table}

\begin{table}[H]
\centering
\caption{Verified literature ephemerides for the 56 planets in category 3.}
\label{tab:verified}
\begin{tabular}{lccl}
Planet & T$_0$ (BJD$_\mathrm{TDB}$) & Period (days) & Reference  \\ [0.3ex] \hline

CoRoT-2b & 2454237.53616 $\pm$ 3e-05 & 1.74299609 $\pm$ 2e-07 & \cite{2016A_A...595A..89B} \\ [0.3ex] 
GJ1214b & 2454934.9177 $\pm$ 3e-05 & 1.58040494 $\pm$ 9e-08 & \cite{2014A_A...565A...7C} \\ [0.3ex] 
GJ436b & 2454865.08403 $\pm$ 3e-05 & 2.643898 $\pm$ 3e-07 & \cite{2014A_A...572A..73L} \\ [0.3ex] 
HAT-P-1b & 2453979.93277 $\pm$ 0.00024 & 4.4652998 $\pm$ 6e-07 & \cite{2014MNRAS.437...46N} \\ [0.3ex] 
HAT-P-3b & 2454856.702 $\pm$ 0.0001 & 2.899736 $\pm$ 2e-06 & \cite{2011AJ....141..179C} \\ [0.3ex] 
HAT-P-4b & 2454245.8152 $\pm$ 0.0002 & 3.0565254 $\pm$ 1.2e-06 & \cite{2012PASP..124..212S} \\ [0.3ex] 
HAT-P-5b & 2455432.4551 $\pm$ 0.0001 & 2.7884736 $\pm$ 5e-07 & \cite{2012MNRAS.422.3099S} \\ [0.3ex] 
HAT-P-9b & 2455484.9131 $\pm$ 0.0004 & 3.9228107 $\pm$ 1e-06 & \cite{2019AJ....157...82W} \\ [0.3ex] 
HAT-P-12b & 2454419.19584 $\pm$ 9e-05 & 3.2130589 $\pm$ 3e-07 & \cite{2016PASP..128b4402S} \\ [0.3ex] 
HAT-P-13b & 2455176.53893 $\pm$ 0.00022 & 2.9162433 $\pm$ 1.2e-06 & \cite{2016PASP..128b4402S} \\ [0.3ex] 
HAT-P-16b & 2455027.59292 $\pm$ 0.00019 & 2.7759704 $\pm$ 7e-07 & \cite{2016PASP..128b4402S} \\ [0.3ex] 
HAT-P-19b & 2455091.535 $\pm$ 0.00015 & 4.0087842 $\pm$ 7e-07 & \cite{2015MNRAS.451.4060S} \\ [0.3ex] 
HAT-P-23b & 2454852.26548 $\pm$ 0.00017 & 1.2128867 $\pm$ 2e-07 & \cite{2016PASP..128b4402S} \\ [0.3ex] 
HAT-P-25b & 2457006.91299 $\pm$ 0.00021 & 3.6528159 $\pm$ 7e-07 & \cite{2019A_A...622A..81M} \\ [0.3ex] 
HAT-P-32b & 2454420.44712 $\pm$ 9e-05 & 2.150008 $\pm$ 1e-06 & \cite{2011ApJ...742...59H} \\ [0.3ex] 
HAT-P-36b & 2455565.1816 $\pm$ 0.0004 & 1.3273468 $\pm$ 5e-07 & \cite{2015A_A...579A.136M} \\ [0.3ex] 
HAT-P-44b & 2456204.47794 $\pm$ 0.00019 & 4.3011886 $\pm$ 1e-06 & \cite{2019A_A...622A..81M} \\ [0.3ex] 
HAT-P-51b & 2456194.1228 $\pm$ 0.0004 & 4.218028 $\pm$ 6e-06 & \cite{2015AJ....150..168H} \\ [0.3ex] 
HAT-P-52b & 2456645.1398 $\pm$ 0.0003 & 2.7535965 $\pm$ 1.1e-06 & \cite{2019A_A...622A..81M} \\ [0.3ex] 
HD189733b & 2454279.437468 $\pm$ 1.5e-05 & 2.21857567 $\pm$ 1.5e-07 & \cite{2010ApJ...721.1861A} \\ [0.3ex] 
KELT-1b & 2457306.976 $\pm$ 0.0003 & 1.2174928 $\pm$ 6e-07 & \cite{2019AJ....158..166B} \\ [0.3ex] 
KELT-3b & 2456269.4899 $\pm$ 0.0003 & 2.703385 $\pm$ 1.8e-06 & \cite{2019A_A...622A..81M} \\ [0.3ex] 
Kepler-6b & 2454954.48742 $\pm$ 3e-05 & 3.23469918 $\pm$ 3e-08 & \cite{2016ApJS..225....9H} \\ [0.3ex] 
Qatar-1b & 2456234.10322 $\pm$ 6e-05 & 1.4200242 $\pm$ 2.2e-07 & \cite{2017AJ....153...78C} \\ [0.3ex] 
Qatar-2b & 2457250.201605 $\pm$ 8e-06 & 1.33711677 $\pm$ 1e-07 & \cite{2017MNRAS.471..394M} \\ [0.3ex] 
TrES-1b & 2453186.807 $\pm$ 0.00012 & 3.030072 $\pm$ 3e-07 & \cite{2012PASP..124..212S} \\ [0.3ex] 
TrES-2b & 2454955.764103 $\pm$ 6e-06 & 2.47061338 $\pm$ 1e-08 & \cite{2016ApJS..225....9H} \\ [0.3ex] 
TrES-3b & 2454185.91116 $\pm$ 6e-05 & 1.30618619 $\pm$ 1.5e-07 & \cite{2013AJ....145...68J} \\ [0.3ex] 
TrES-4b & 2454230.9056 $\pm$ 0.0003 & 3.5539277 $\pm$ 5e-07 & \cite{2015A_A...575L..15S} \\ [0.3ex] 
TrES-5b & 2455443.25271 $\pm$ 0.00011 & 1.48224754 $\pm$ 1.7e-07 & \cite{2016AcA....66...55M} \\ [0.3ex] 
WASP-1b & 2453912.5151 $\pm$ 0.0003 & 2.5199454 $\pm$ 5e-07 & \cite{2014AcA....64...27M} \\ [0.3ex] 
WASP-2b & 2453991.51536 $\pm$ 0.00018 & 2.1522213 $\pm$ 4e-07 & \cite{2012PASP..124..212S} \\ [0.3ex] 
WASP-10b & 2454664.03804 $\pm$ 6e-05 & 3.0927295 $\pm$ 3e-07 & \cite{2016PASP..128b4402S} \\ [0.3ex] 
WASP-11b & 2454729.9072 $\pm$ 0.0002 & 3.7224797 $\pm$ 4e-07 & \cite{2015A_A...579A.136M} \\ [0.3ex] 
WASP-12b & 2456176.66826 $\pm$ 8e-05 & 1.0914203 $\pm$ 1.4e-07 & \cite{2017AJ....153...78C} \\ [0.3ex] 
WASP-19b & 2455708.534626 $\pm$ 1.9e-05 & 0.78883899 $\pm$ 4e-08 & \cite{2016ApJ...823..122W} \\ [0.3ex] 
WASP-24b & 2455687.75616 $\pm$ 0.00016 & 2.3412217 $\pm$ 8e-07 & \cite{2014MNRAS.444..776S} \\ [0.3ex] 
WASP-36b & 2455569.8377 $\pm$ 0.0005 & 1.53736596 $\pm$ 2.4e-07 & \cite{2016MNRAS.459.1393M} \\ [0.3ex] 
WASP-43b & 2455528.86863 $\pm$ 5e-05 & 0.81347398 $\pm$ 4e-08 & \cite{2016AJ....151..137H} \\ [0.3ex] 
WASP-48b & 2455364.55241 $\pm$ 0.00024 & 2.1436354 $\pm$ 6e-07 & \cite{2015A_A...577A..54C} \\ [0.3ex] 
WASP-49b & 2456267.6839 $\pm$ 0.00013 & 2.7817362 $\pm$ 1.4e-06 & \cite{2017A_A...602A..36W} \\ [0.3ex] 
WASP-50b & 2455558.6124 $\pm$ 0.0002 & 1.9550938 $\pm$ 1.3e-06 & \cite{2013MNRAS.431..966T} \\ [0.3ex] 
WASP-52b & 2456862.79795 $\pm$ 7e-05 & 1.74978114 $\pm$ 1.4e-07 & \cite{2019MNRAS.486.2290O} \\ [0.3ex] 
WASP-57b & 2456058.5491 $\pm$ 0.00016 & 2.8389186 $\pm$ 8e-07 & \cite{2015MNRAS.454.3094S} \\ [0.3ex] 
WASP-65b & 2456110.68772 $\pm$ 0.00015 & 2.3114243 $\pm$ 1.5e-06 & \cite{2013A_A...559A..36G} \\ [0.3ex] 
WASP-80b & 2456487.42578 $\pm$ 3e-05 & 3.0678523 $\pm$ 8e-07 & \cite{2015MNRAS.450.2279T} \\ [0.3ex] 
WASP-85Ab & 2456847.473634 $\pm$ 1.4e-05 & 2.6556777 $\pm$ 4e-07 & \cite{2016AJ....151..150M} \\ [0.3ex] 
WASP-92b & 2456381.2842 $\pm$ 0.0003 & 2.1746742 $\pm$ 1.6e-06 & \cite{2016MNRAS.463.3276H} \\ [0.3ex] 
WASP-93b & 2456079.565 $\pm$ 0.0004 & 2.7325321 $\pm$ 2e-06 & \cite{2016MNRAS.463.3276H} \\ [0.3ex] 
WASP-103b & 2456836.29644 $\pm$ 6e-05 & 0.9255456 $\pm$ 1.3e-06 & \cite{2015MNRAS.447..711S} \\ [0.3ex] 
WASP-113b & 2457197.09823 $\pm$ 4e-05 & 4.542169 $\pm$ 4e-06 & \cite{2016A_A...593A.113B} \\ [0.3ex] 
WASP-114b & 2456667.73661 $\pm$ 0.00021 & 1.5487743 $\pm$ 1.2e-06 & \cite{2016A_A...593A.113B} \\ [0.3ex] 
WASP-153b & 2453142.543 $\pm$ 0.003 & 3.332609 $\pm$ 2e-06 & \cite{2018A_A...610A..63D} \\ [0.3ex] 
XO-1b & 2453887.74774 $\pm$ 0.00022 & 3.9415069 $\pm$ 1.8e-06 & \cite{2010ApJ...719.1796B} \\ [0.3ex] 
XO-2Nb & 2455565.54648 $\pm$ 5e-05 & 2.6158592 $\pm$ 3e-07 & \cite{2015A_A...575A.111D} \\ [0.3ex] 
XO-5b & 2456864.3137 $\pm$ 0.0002 & 4.1877558 $\pm$ 6e-07 & \cite{2015AcA....65..117S} \\ [0.3ex] 

\end{tabular}
\end{table}

\begin{table}[H]
\centering
\caption{Sources for the parameters used in the analysis of all 120 planets in this data release. For each planet, three  references are provided, one for the stellar parameters used to calculate the limb darkening, one for the transit parameters and one for the initial ephemeris (before any update).}
\label{tab:ref}
\begin{tabular}{llllll}

CoRoT-2b& \cite{2010A_A...517A..40C} \cite{2008A_A...482L..21A} \cite{2016A_A...595A..89B} & HAT-P-65b& \cite{2016AJ....152..182H} \cite{2016AJ....152..182H} \cite{2016AJ....152..182H} & WASP-26b& \cite{2010A_A...520A..56S} \cite{2014MNRAS.444..776S} \cite{2014MNRAS.444..776S} \\ [0.3ex] 
GJ1214b& \cite{2009Natur.462..891C} \cite{2012ApJ...747...35B} \cite{2014A_A...565A...7C} & HATS-6b& \cite{2015AJ....149..166H} \cite{2015AJ....149..166H} \cite{2015AJ....149..166H} & WASP-31b& \cite{2011A_A...531A..60A} \cite{2011A_A...531A..60A} \cite{2011A_A...531A..60A} \\ [0.3ex] 
GJ436b& \cite{2008ApJ...677.1324T} \cite{2011ApJ...735...27K} \cite{2014A_A...572A..73L} & HATS-22b& \cite{2017MNRAS.468..835B} \cite{2017MNRAS.468..835B} \cite{2017MNRAS.468..835B} & WASP-32b& \cite{2010PASP..122.1465M} \cite{2010PASP..122.1465M} \cite{2012PASP..124..212S} \\ [0.3ex] 
HAT-P-1b& \cite{2008ApJ...677.1324T} \cite{2014MNRAS.437...46N} \cite{2014MNRAS.437...46N} & HATS-25b& \cite{2016AJ....152..108E} \cite{2016AJ....152..108E} \cite{2016AJ....152..108E} & WASP-36b& \cite{2012AJ....143...81S} \cite{2016MNRAS.459.1393M} \cite{2016MNRAS.459.1393M} \\ [0.3ex] 
HAT-P-3b& \cite{2008ApJ...677.1324T} \cite{2011AJ....141..179C} \cite{2011AJ....141..179C} & HD189733b& \cite{2008ApJ...677.1324T} \cite{2014ApJ...786...22M} \cite{2010ApJ...721.1861A} & WASP-43b& \cite{2011A_A...535L...7H} \cite{2016AJ....151..137H} \cite{2016AJ....151..137H} \\ [0.3ex] 
HAT-P-4b& \cite{2008ApJ...677.1324T} \cite{2011ApJ...726...94C} \cite{2012PASP..124..212S} & HD209458b& \cite{2008ApJ...677.1324T} \cite{2008ApJ...677.1324T} \cite{2007ApJ...655..564K} & WASP-48b& \cite{2011AJ....142...86E} \cite{2015A_A...577A..54C} \cite{2015A_A...577A..54C} \\ [0.3ex] 
HAT-P-5b& \cite{2008ApJ...677.1324T} \cite{2008ApJ...677.1324T} \cite{2012MNRAS.422.3099S} & HD80606b& \cite{2013A_A...558A.106M} \cite{2010A_A...516A..95H} \cite{2010A_A...516A..95H} & WASP-49b& \cite{2012A_A...544A..72L} \cite{2017A_A...602A..36W} \cite{2017A_A...602A..36W} \\ [0.3ex] 
HAT-P-6b& \cite{2008ApJ...673L..79N} \cite{2008ApJ...677.1324T} \cite{2008ApJ...673L..79N} & K2-29b& \cite{2016ApJ...824...55S} \cite{2016ApJ...824...55S} \cite{2016ApJ...824...55S} & WASP-50b& \cite{2011A_A...533A..88G} \cite{2011A_A...533A..88G} \cite{2013MNRAS.431..966T} \\ [0.3ex] 
HAT-P-7b& \cite{2008ApJ...680.1450P} \cite{2016ApJ...823..122W} \cite{2016ApJS..225....9H} & K2-30b& \cite{2016AJ....151..171J} \cite{2016AJ....151..171J} \cite{2016AJ....151..171J} & WASP-52b& \cite{2013A_A...549A.134H} \cite{2013A_A...549A.134H} \cite{2019MNRAS.486.2290O} \\ [0.3ex] 
HAT-P-8b& \cite{2009ApJ...704.1107L} \cite{2013A_A...551A..11M} \cite{2013A_A...551A..11M} & KELT-1b& \cite{2012ApJ...761..123S} \cite{2019AJ....158..166B} \cite{2019AJ....158..166B} & WASP-55b& \cite{2012MNRAS.426..739H} \cite{2016MNRAS.457.4205S} \cite{2016MNRAS.457.4205S} \\ [0.3ex] 
HAT-P-9b& \cite{2009ApJ...690.1393S} \cite{2019AJ....157...82W} \cite{2019AJ....157...82W} & KELT-3b& \cite{2013ApJ...773...64P} \cite{2013ApJ...773...64P} \cite{2019A_A...622A..81M} & WASP-57b& \cite{2013A_A...551A..73F} \cite{2015MNRAS.454.3094S} \cite{2015MNRAS.454.3094S} \\ [0.3ex] 
HAT-P-12b& \cite{2009ApJ...706..785H} \cite{2009ApJ...706..785H} \cite{2016PASP..128b4402S} & KELT-7b& \cite{2015AJ....150...12B} \cite{2015AJ....150...12B} \cite{2015AJ....150...12B} & WASP-58b& \cite{2013A_A...549A.134H} \cite{2013A_A...549A.134H} \cite{2019A_A...622A..81M} \\ [0.3ex] 
HAT-P-13b& \cite{2009ApJ...707..446B} \cite{2009ApJ...707..446B} \cite{2016PASP..128b4402S} & KELT-8b& \cite{2015ApJ...810...30F} \cite{2015ApJ...810...30F} \cite{2019A_A...622A..81M} & WASP-65b& \cite{2013A_A...559A..36G} \cite{2013A_A...559A..36G} \cite{2013A_A...559A..36G} \\ [0.3ex] 
HAT-P-16b& \cite{2010ApJ...720.1118B} \cite{2010ApJ...720.1118B} \cite{2016PASP..128b4402S} & KELT-15b& \cite{2016AJ....151..138R} \cite{2016AJ....151..138R} \cite{2016AJ....151..138R} & WASP-67b& \cite{2012MNRAS.426..739H} \cite{2012MNRAS.426..739H} \cite{2012MNRAS.426..739H} \\ [0.3ex] 
HAT-P-17b& \cite{2012ApJ...749..134H} \cite{2012ApJ...749..134H} \cite{2012ApJ...749..134H} & KELT-16b& \cite{2017AJ....153...97O} \cite{2017AJ....153...97O} \cite{2017AJ....153...97O} & WASP-69b& \cite{2014MNRAS.445.1114A} \cite{2014MNRAS.445.1114A} \cite{2014MNRAS.445.1114A} \\ [0.3ex] 
HAT-P-18b& \cite{2011ApJ...726...52H} \cite{2017MNRAS.468.3907K} \cite{2015MNRAS.451.4060S} & KPS-1b& \cite{2018PASP..130g4401B} \cite{2018PASP..130g4401B} \cite{2018PASP..130g4401B} & WASP-74b& \cite{2015AJ....150...18H} \cite{2015AJ....150...18H} \cite{2015AJ....150...18H} \\ [0.3ex] 
HAT-P-19b& \cite{2011ApJ...726...52H} \cite{2011ApJ...726...52H} \cite{2015MNRAS.451.4060S} & Kepler-6b& \cite{2010ApJ...713L.136D} \cite{2015ApJ...804..150E} \cite{2016ApJS..225....9H} & WASP-75b& \cite{2013A_A...559A..36G} \cite{2013A_A...559A..36G} \cite{2013A_A...559A..36G} \\ [0.3ex] 
HAT-P-20b& \cite{2011ApJ...742..116B} \cite{2011ApJ...742..116B} \cite{2011ApJ...742..116B} & Qatar-1b& \cite{2011MNRAS.417..709A} \cite{2017AJ....153...78C} \cite{2017AJ....153...78C} & WASP-76b& \cite{2016A_A...585A.126W} \cite{2016A_A...585A.126W} \cite{2016A_A...585A.126W} \\ [0.3ex] 
HAT-P-22b& \cite{2011ApJ...742..116B} \cite{2011ApJ...742..116B} \cite{2011ApJ...742..116B} & Qatar-2b& \cite{2012ApJ...750...84B} \cite{2017MNRAS.471..394M} \cite{2017MNRAS.471..394M} & WASP-77Ab& \cite{2013PASP..125...48M} \cite{2013PASP..125...48M} \cite{2013PASP..125...48M} \\ [0.3ex] 
HAT-P-23b& \cite{2011ApJ...742..116B} \cite{2015A_A...577A..54C} \cite{2016PASP..128b4402S} & Qatar-3b& \cite{2017AJ....153..200A} \cite{2017AJ....153..200A} \cite{2019A_A...622A..81M} & WASP-80b& \cite{2013A_A...551A..80T} \cite{2015MNRAS.450.2279T} \cite{2015MNRAS.450.2279T} \\ [0.3ex] 
HAT-P-24b& \cite{2010ApJ...725.2017K} \cite{2010ApJ...725.2017K} \cite{2010ApJ...725.2017K} & Qatar-4b& \cite{2017AJ....153..200A} \cite{2017AJ....153..200A} \cite{2019A_A...622A..81M} & WASP-82b& \cite{2016A_A...585A.126W} \cite{2016A_A...585A.126W} \cite{2015AcA....65..117S} \\ [0.3ex] 
HAT-P-25b& \cite{2012ApJ...745...80Q} \cite{2018PASP..130f4401W} \cite{2019A_A...622A..81M} & Qatar-5b& \cite{2017AJ....153..200A} \cite{2017AJ....153..200A} \cite{2019A_A...622A..81M} & WASP-83b& \cite{2015AJ....150...18H} \cite{2015AJ....150...18H} \cite{2015AJ....150...18H} \\ [0.3ex] 
HAT-P-26b& \cite{2011ApJ...728..138H} \cite{2011ApJ...728..138H} \cite{2016ApJ...817..141S} & TrES-1b& \cite{2008ApJ...677.1324T} \cite{2008ApJ...677.1324T} \cite{2012PASP..124..212S} & WASP-84b& \cite{2014MNRAS.445.1114A} \cite{2014MNRAS.445.1114A} \cite{2014MNRAS.445.1114A} \\ [0.3ex] 
HAT-P-28b& \cite{2011ApJ...733..116B} \cite{2011ApJ...733..116B} \cite{2011ApJ...733..116B} & TrES-2b& \cite{2008ApJ...677.1324T} \cite{2015ApJ...804..150E} \cite{2016ApJS..225....9H} & WASP-85Ab& \cite{2014arXiv1412.7761B} \cite{2014arXiv1412.7761B} \cite{2016AJ....151..150M} \\ [0.3ex] 
HAT-P-30b& \cite{2011ApJ...735...24J} \cite{2016AcA....66...55M} \cite{2016AcA....66...55M} & TrES-3b& \cite{2008ApJ...677.1324T} \cite{2011ApJ...726...94C} \cite{2013AJ....145...68J} & WASP-92b& \cite{2016MNRAS.463.3276H} \cite{2016MNRAS.463.3276H} \cite{2016MNRAS.463.3276H} \\ [0.3ex] 
HAT-P-32b& \cite{2011ApJ...742...59H} \cite{2019AJ....157...82W} \cite{2011ApJ...742...59H} & TrES-4b& \cite{2008ApJ...677.1324T} \cite{2015A_A...575L..15S} \cite{2015A_A...575L..15S} & WASP-93b& \cite{2016MNRAS.463.3276H} \cite{2016MNRAS.463.3276H} \cite{2016MNRAS.463.3276H} \\ [0.3ex] 
HAT-P-36b& \cite{2012AJ....144...19B} \cite{2019AJ....157...82W} \cite{2015A_A...579A.136M} & TrES-5b& \cite{2011ApJ...741..114M} \cite{2016AcA....66...55M} \cite{2016AcA....66...55M} & WASP-103b& \cite{2014A_A...562L...3G} \cite{2016MNRAS.463...37S} \cite{2015MNRAS.447..711S} \\ [0.3ex] 
HAT-P-37b& \cite{2012AJ....144...19B} \cite{2016AcA....66...55M} \cite{2016AcA....66...55M} & WASP-1b& \cite{2008ApJ...677.1324T} \cite{2014AcA....64...27M} \cite{2014AcA....64...27M} & WASP-104b& \cite{2014A_A...570A..64S} \cite{2014A_A...570A..64S} \cite{2014A_A...570A..64S} \\ [0.3ex] 
HAT-P-38b& \cite{2012PASJ...64...97S} \cite{2012PASJ...64...97S} \cite{2019A_A...622A..81M} & WASP-2b& \cite{2008ApJ...677.1324T} \cite{2010MNRAS.408.1680S} \cite{2012PASP..124..212S} & WASP-107b& \cite{2017A_A...604A.110A} \cite{2017MNRAS.469.1622M} \cite{2017MNRAS.469.1622M} \\ [0.3ex] 
HAT-P-39b& \cite{2012AJ....144..139H} \cite{2012AJ....144..139H} \cite{2012AJ....144..139H} & WASP-3b& \cite{2008MNRAS.385.1576P} \cite{2011ApJ...726...94C} \cite{2011ApJ...726...94C} & WASP-113b& \cite{2016A_A...593A.113B} \cite{2016A_A...593A.113B} \cite{2016A_A...593A.113B} \\ [0.3ex] 
HAT-P-41b& \cite{2012AJ....144..139H} \cite{2012AJ....144..139H} \cite{2012AJ....144..139H} & WASP-10b& \cite{2009MNRAS.392.1585C} \cite{2009ApJ...692L.100J} \cite{2016PASP..128b4402S} & WASP-114b& \cite{2016A_A...593A.113B} \cite{2016A_A...593A.113B} \cite{2016A_A...593A.113B} \\ [0.3ex] 
HAT-P-44b& \cite{2014AJ....147..128H} \cite{2014AJ....147..128H} \cite{2019A_A...622A..81M} & WASP-11b& \cite{2009A_A...502..395W} \cite{2014AJ....147...92W} \cite{2015A_A...579A.136M} & WASP-127b& \cite{2017A_A...599A...3L} \cite{2017A_A...602L..15P} \cite{2017A_A...599A...3L} \\ [0.3ex] 
HAT-P-46b& \cite{2014AJ....147..128H} \cite{2014AJ....147..128H} \cite{2019A_A...622A..81M} & WASP-12b& \cite{2009ApJ...693.1920H} \cite{2017AJ....153...78C} \cite{2017AJ....153...78C} & WASP-153b& \cite{2018A_A...610A..63D} \cite{2018A_A...610A..63D} \cite{2018A_A...610A..63D} \\ [0.3ex] 
HAT-P-49b& \cite{2014AJ....147...84B} \cite{2014AJ....147...84B} \cite{2014AJ....147...84B} & WASP-13b& \cite{2009A_A...502..391S} \cite{2009A_A...502..391S} \cite{2012MNRAS.419.1248B} & WASP-167b& \cite{2017MNRAS.471.2743T} \cite{2017MNRAS.471.2743T} \cite{2017MNRAS.471.2743T} \\ [0.3ex] 
HAT-P-51b& \cite{2015AJ....150..168H} \cite{2015AJ....150..168H} \cite{2015AJ....150..168H} & WASP-15b& \cite{2009AJ....137.4834W} \cite{2013MNRAS.434.1300S} \cite{2013MNRAS.434.1300S} & XO-1b& \cite{2008ApJ...677.1324T} \cite{2008ApJ...677.1324T} \cite{2010ApJ...719.1796B} \\ [0.3ex] 
HAT-P-52b& \cite{2015AJ....150..168H} \cite{2015AJ....150..168H} \cite{2019A_A...622A..81M} & WASP-16b& \cite{2009ApJ...703..752L} \cite{2013MNRAS.434.1300S} \cite{2013MNRAS.434.1300S} & XO-2Nb& \cite{2008ApJ...677.1324T} \cite{2015A_A...575A.111D} \cite{2015A_A...575A.111D} \\ [0.3ex] 
HAT-P-53b& \cite{2015AJ....150..168H} \cite{2015AJ....150..168H} \cite{2015AJ....150..168H} & WASP-19b& \cite{2010ApJ...708..224H} \cite{2016ApJ...823..122W} \cite{2016ApJ...823..122W} & XO-3b& \cite{2008ApJ...677..657J} \cite{2014ApJ...794..134W} \cite{2014ApJ...794..134W} \\ [0.3ex] 
HAT-P-54b& \cite{2015AJ....149..149B} \cite{2015AJ....149..149B} \cite{2015AJ....149..149B} & WASP-21b& \cite{2010A_A...519A..98B} \cite{2010A_A...519A..98B} \cite{2015MNRAS.451.4060S} & XO-4b& \cite{2008arXiv0805.2921M} \cite{2010PASJ...62L..61N} \cite{2010PASJ...62L..61N} \\ [0.3ex] 
HAT-P-56b& \cite{2015AJ....150...85H} \cite{2015AJ....150...85H} \cite{2015AJ....150...85H} & WASP-24b& \cite{2010ApJ...720..337S} \cite{2014MNRAS.444..776S} \cite{2014MNRAS.444..776S} & XO-5b& \cite{2008ApJ...686.1331B} \cite{2015AcA....65..117S} \cite{2015AcA....65..117S} \\ [0.3ex] 
HAT-P-57b& \cite{2015AJ....150..197H} \cite{2015AJ....150..197H} \cite{2015AJ....150..197H} & WASP-25b& \cite{2011MNRAS.410.1631E} \cite{2014MNRAS.444..776S} \cite{2014MNRAS.444..776S} & XO-6b& \cite{2017AJ....153...94C} \cite{2017AJ....153...94C} \cite{2017AJ....153...94C} \\ [0.3ex]

\end{tabular}
\end{table}

\newpage

\bibliographystyle{spphys}   
\bibliography{references,references_a,ref_ephem}   

\end{document}